\RequirePackage{fix-cm}
\documentclass[smallextended]{svjour3}       
\smartqed  
\usepackage{tablefootnote}
\usepackage{graphicx}
\usepackage{multirow}
\usepackage{algpseudocode} 
\usepackage{subfig}
\usepackage[dvipsnames]{xcolor}
\usepackage[ruled,vlined]{algorithm2e}
\usepackage{url}
\usepackage{adjustbox}
\usepackage[authoryear,longnamesfirst]{natbib}
\graphicspath{ {Figures/} }
\begin{document}

\title{Gender imbalance in retracted publications is more favorable toward women authors}
\author{Kiran Sharma         \and
        Harsha V.  Garine   \and
       Satyam Mukherjee
}


\institute{
           K. Sharma \& H.  V. Garine \at
              School of Engineering and Technology, BML Munjal University, Gurugram, Haryana-122413, India \\
              \email{kiran.sharma@bmu.edu.in}       
                 \and
             S. Mukharjee \at
             School of Management and Entrepreneurship, Shiv Nadar Institution of Eminence,  Greater Noida, Uttar Pradesh-201314, India  \\
             \email{satyam.mukherjee@snu.edu.in }           
}

%
%
%


\date{Received: date / Accepted: date}

\maketitle
\begin{abstract}

Numerous studies in the literature highlight that women are underrepresented in the scientific domain which further leads to the under-representation of women in prestigious publications, authorship positions, and collaboration.  However, the representation of women in scientific misconduct has not been studied yet, hence the study aims to investigate the female involvement and authorship position in retracted publications. To conduct the analysis, 3750 retracted scientific papers were extracted from the Web of Science and the respective gender was identified for each author. The evaluation included the year-wise representation of female authors, females at various authorship positions, collaboration,  and female-to-male odds ratio. In all,  26.43\% of authorship is held by women and the share of male-female collaborative retracted publications is 55.11\%. In retracted publications, women are less likely to hold the last authorship and more likely to hold the middle authorship position.


\end{abstract}
\section{Introduction}

Extant literature in STEM suggests gender disparity act as a deterrent for women in the scientific workforce~\citep{vasarhelyi2021gender, fang2013males,aiston2016women}. The existing ``glass ceiling'' prohibits women from receiving scientific recognition, delays hiring and promotion, and receives fewer citations compared to their male counterparts~\citep{vasarhelyi2021gender}. Recent research involving the study of scientific teams suggests men benefit more from collaborative work compared to women. Further, a recent study on gender disparities in recognition of work observed higher online visibility for men and negligible online success for women. We build upon prior research on gender disparities and investigate gender imbalance in retracted publications.

Recent studies involving retracted publications suggest an increasing trend in this scientific misconduct~\citep{steen2011retractions, fang2012misconduct, steen2013has, azoulay2017career}. Irrespective of the reasons behind the increasing trend, retracted publications incur a loss of reputation in the scientific communities~\citep{van2011science, peng2022dynamics}. In this work, we combine the two streams of research on gender disparities and retraction trends. Leveraging data from retracted publications and identifying the gender of authors of retracted papers we investigate whether gender disparity hurts women scholars more compared to their male counterparts.

The rise of teams in scientific research has witnessed greater involvement of women scholars in recent times~\citep{agogino2007beyond, yang2022gender}. Even though prior research suggests a negligible benefit of women in collaborative settings, it is unclear how women authors are affected during retractions of publications~\citep{sharma2021team}. In terms of online scholarly activities, it has been observed women tend to blog less and avoid self-promotion. The question thus arises whether women are as likely as men to produce erroneous publications. Further, what authorship role do women have in such retracted publications?
Our research objective is two-fold. First, we investigate the female involvement in the retracted publications. Next, we evaluate the proportion of female authorship position and their collaboration in retracted publications.

The study is organized as follows: Section~\ref{method} is on data collection and filtration. Results are explained in Section~\ref{results}. Finally, the conclusion is presented in Section~\ref{discussion}.

\section{Methodology}
In this section we will provide overview about the data selection,  data filtration, and gender identification process.
\label{method}
\subsection{Data acquisition and filtration}

We performed a Web of Science (WoS) search for the period from 1989 to 2021 to retrieve all articles listed as ``Retracted". A total of 5845 research papers were retracted from the portal. These papers were published in more than 1897 journals. The data has all the necessary meta-data for the analysis which includes author information, publisher details, year of publication, citation count, and discipline of research. The papers retracted before 2006 did not provide the complete author's details in the database, hence we removed those papers from our study and left with 3750 papers. The authors' names and respective positions from 3750 papers were extracted to get the information of 30412 authors. Further, the authors with initials in their names were removed, and 19164 authors were filtered. For each entry, the data set was prepared on the following variables: manuscript title, year of retraction, first, middle,  and last names of the authors, journal impact factor,  country of provenance of all authors, and discipline of publication.

\subsection{Gender determination}
To determine the author's gender, an online database named ``Gender API'' was used. This third-party Gender API determines gender by first name and country. This API has been used by many researchers prior in their work to examine the gender disparity in the authorship of academic articles. The gender extraction was performed based on the author's first name and the country for the accuracy of search results. The gender API provides an accuracy score for the prediction of gender across each name from its sampled data. The accuracy score varies from 0 (not confirmed) to 100 (highly confirmed).  The threshold for authors selection was chosen with accuracy scores above 60. Hence, 17634 authors' gender with a given accuracy were filtered. Further, to cross-verify the data, a random sample of 10\% of the data was selected and a manual recheck was performed on the author's gender from their scholar profiles or institute profiles. This completes the gender extraction process.

\begin{figure}[!h]
    \centering
    \includegraphics[width=0.75\linewidth]{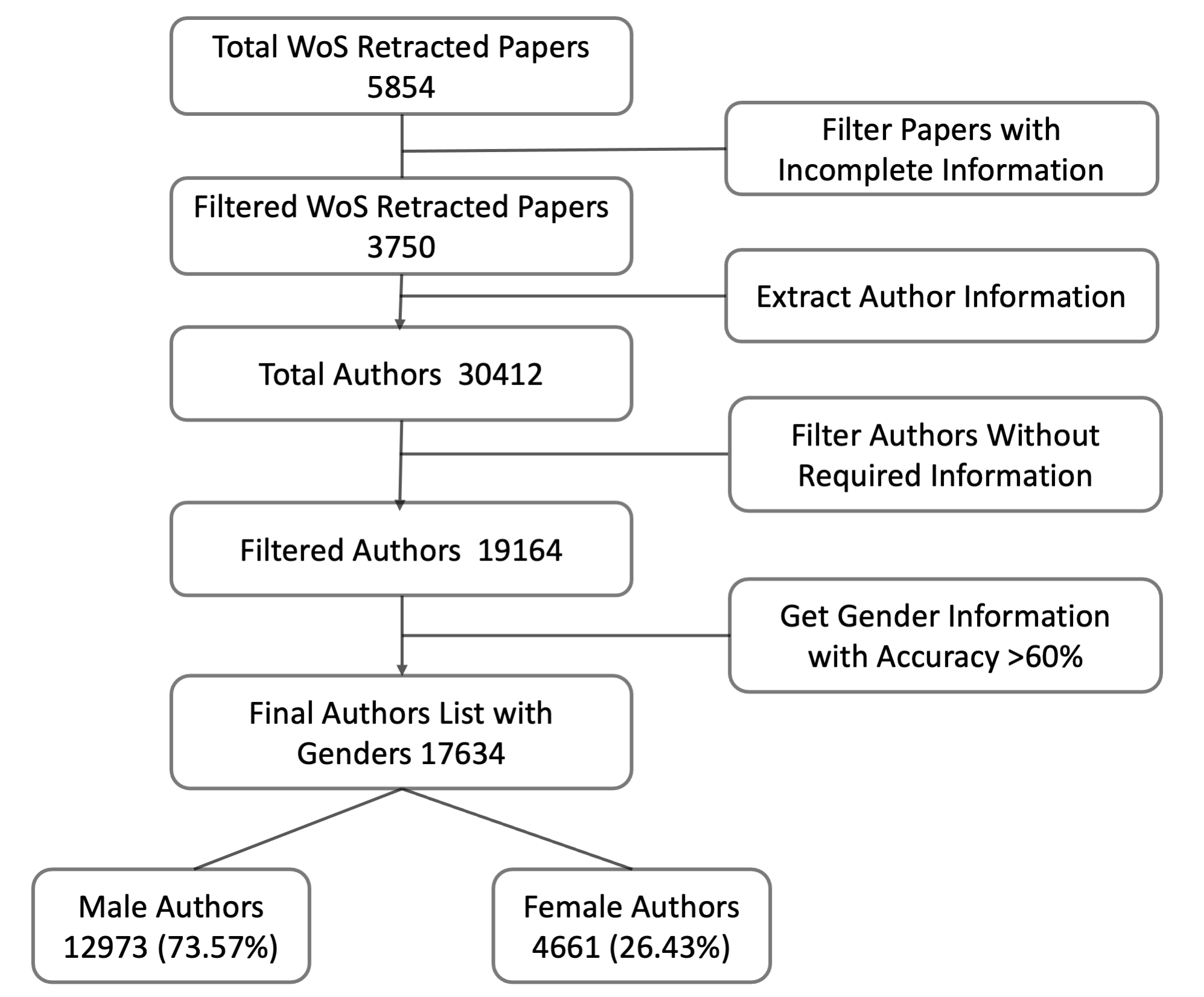}     
    \caption{Flowchart representing the data selection, filtration and gender extraction process.}
 \label{fig:1}
\end{figure}

\section{Results}
\label{results}
\subsection{Female authorship trend and position}

\begin{figure}[!h]
    \centering
    \includegraphics[width=0.98\linewidth]{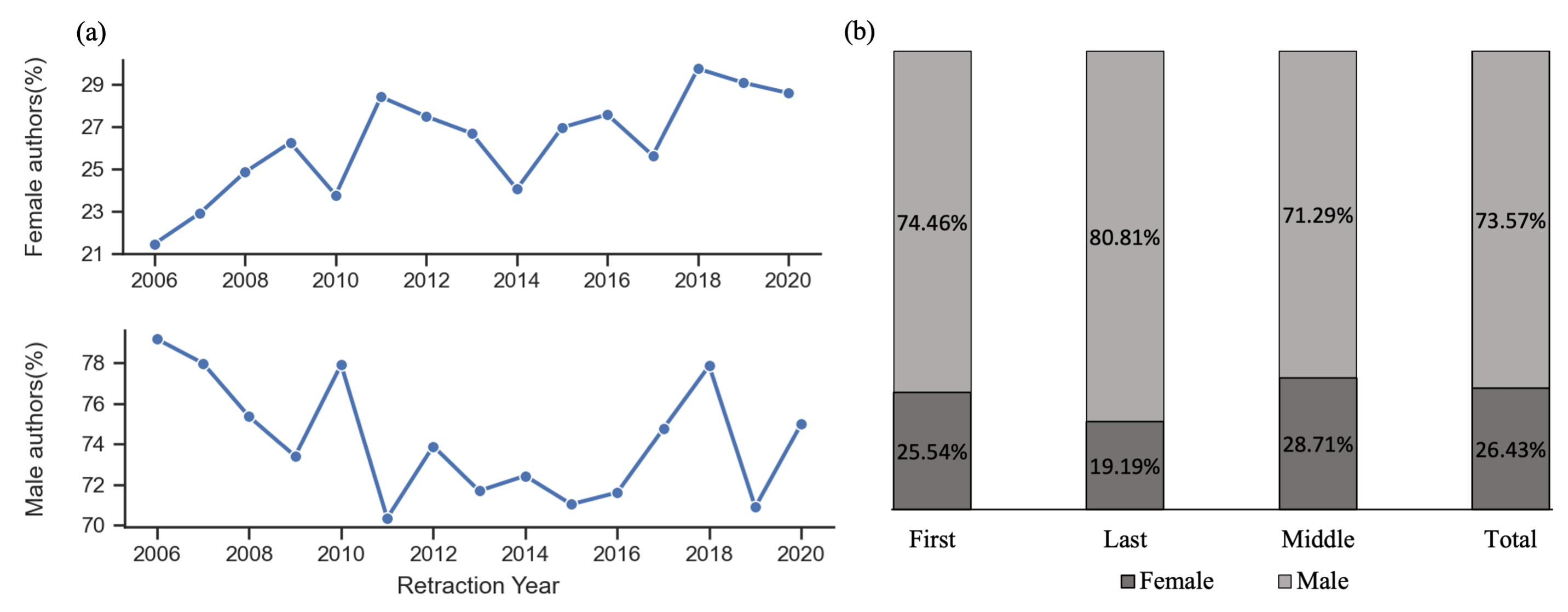}
\caption{(a) Yearwise trend of male and female contribution in retracted publications. (b) The global share of females as first,  last, and middle authorship position. Females comprise 26.43\% of total authorship than males (73.57\%). }
\label{fig:Fig2}   
\end{figure}

Fig.~\ref{fig:Fig2} provides descriptive information on the women's representation (a) year-wise retracted articles from 2006 to 2020, and (b) at different authorship positions as first, last, middle, and total. In year-wise female authorship (Fig.~\ref{fig:Fig2}(a)), the observed trend is a little upward.
In Fig.~\ref{fig:Fig2}(b), women comprise 25.54\% of first authors, 28.71\% of middle authors, and 19.19\% of last authors. In the full author's sample, on average 26.43\% of women's appeared in retracted publications.

\subsection{ Gender specific collaboration and odds ratio}
\begin{figure}[!h]
    \centering
    \includegraphics[width=0.43\linewidth]{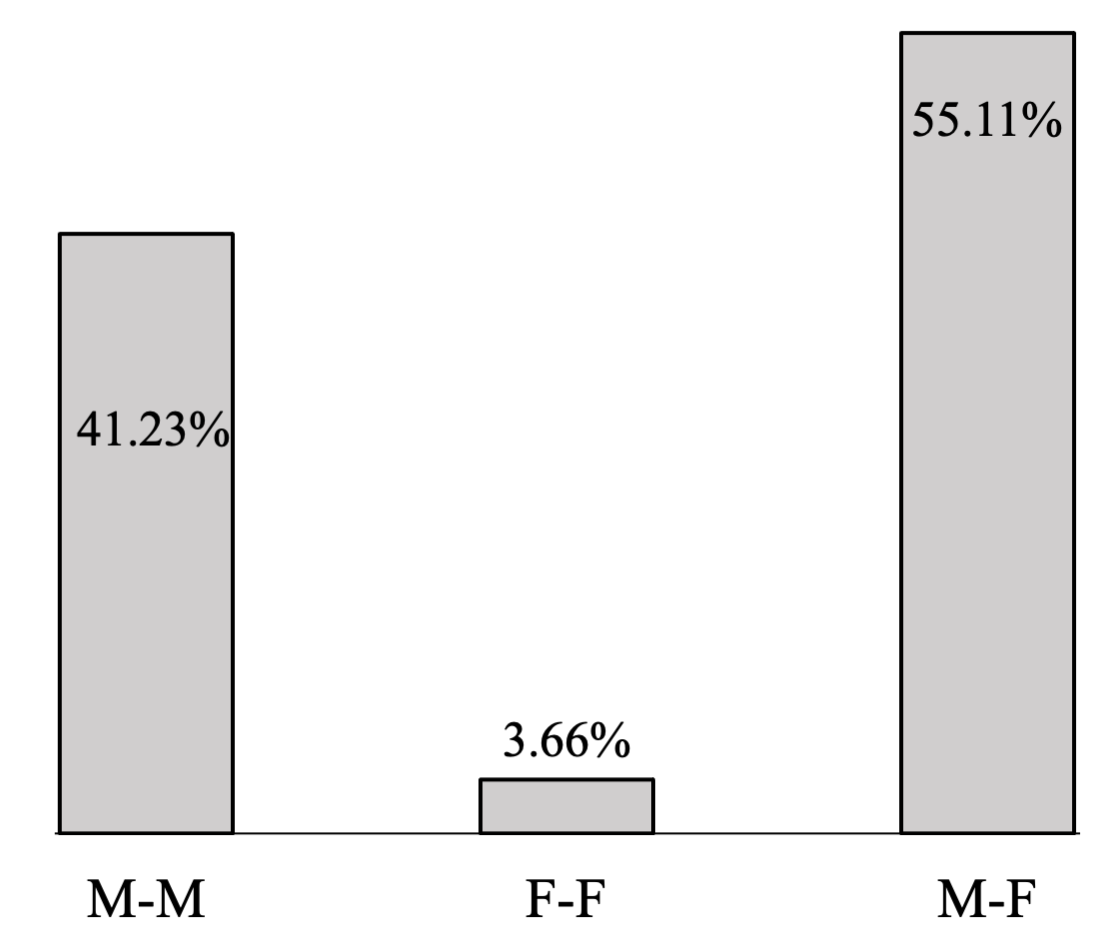}
     \llap{\parbox[b]{2.0in}{(a)\\\rule{0ex}{1.6in}}}
       \includegraphics[width=0.52\linewidth]{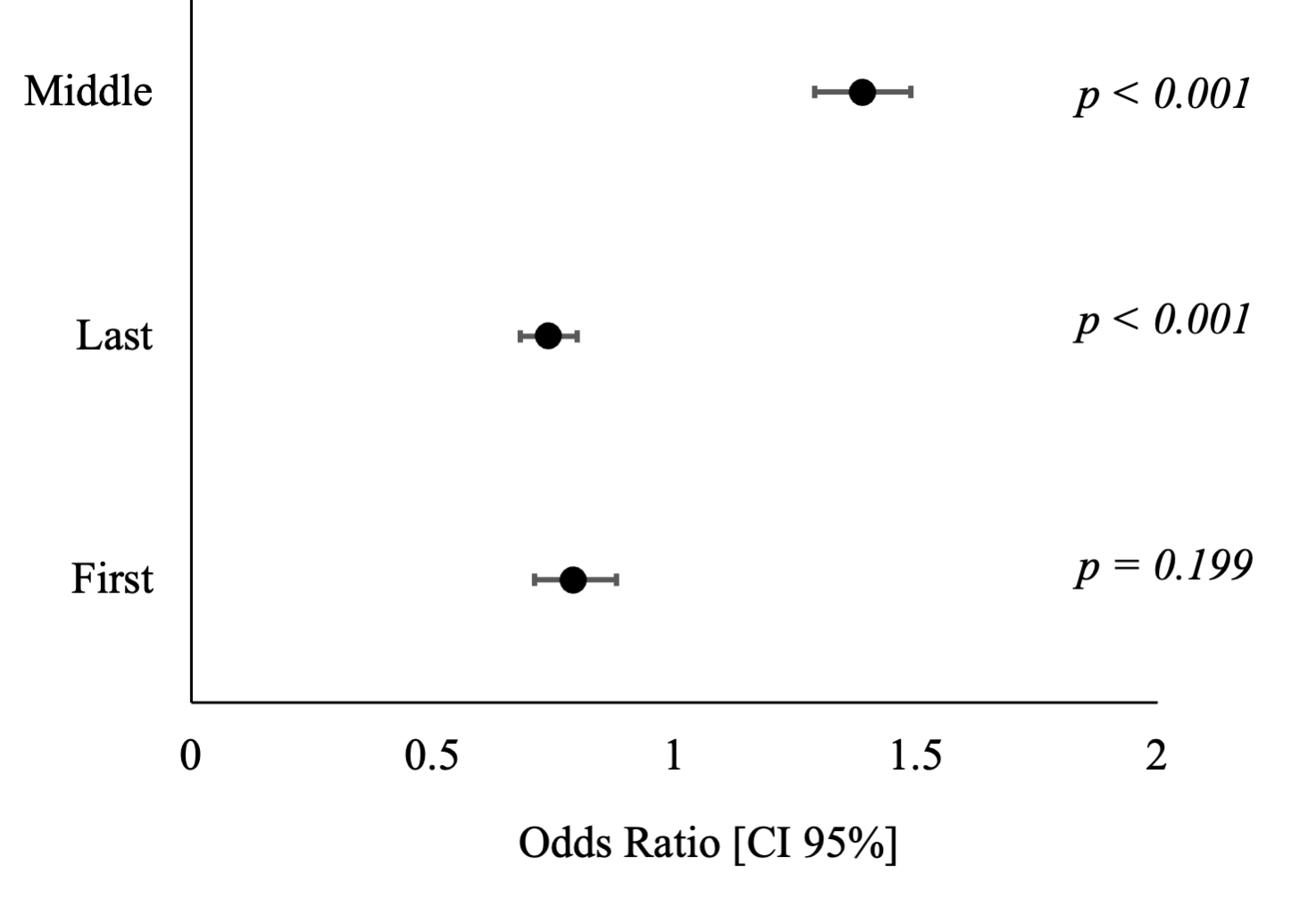}
        \llap{\parbox[b]{2.0in}{(b)\\\rule{0ex}{1.6in}}}
\caption{(a) Gender-specific collaboration in retracted publications where M-M represents all male authors collaboration, F-F represents all female authors collaboration, and M-F represents male-female collaboration. (b) Female to Male odds ratio. Females are less likely to be last and more likely to be the middle author.}
\label{fig:Fig3}   
\end{figure}

In sciences, the first authorship position is usually considered as a junior author who executed the research, while the last authorship position points towards the author who leads and funds the research- the senior author. The male-female collaboration at different levels also points toward the active involvement of women in the research.
To analyze gender-specific collaboration in retracted publications,  papers were categorized as M-M (male-male), F-F (female-female), and M-F(male-female).  41.23\% retractions reported from M-M collaboration, 3.66\% from F-F, and 55.11\% from M-F collaboration (Fig~\ref{fig:Fig3}(a)). Although the retraction reported from the male-female collaboration is higher women are less likely to hold the last authorship position (OR = 0.62,  $p<0.001$) and more likely to hold the middle authorship position (OR= 1.38,  $p<0.001$) (Fig~\ref{fig:Fig3}(b)).

\begin{table}[!h]
\centering
\caption{mlogit estimation.}
\begin{tabular}{|lc|}
\hline
\multicolumn{1}{|l|}{} & \multicolumn{1}{l|}{\begin{tabular}[c]{@{}l@{}}Base category\\ (First author female)\end{tabular}} \\ \hline
\multicolumn{2}{|l|}{\textit{\begin{tabular}[c]{@{}l@{}}Independent variable\\ (Female dummy)\end{tabular}}}                \\ \hline
\multicolumn{1}{|l|}{Last author}            & \begin{tabular}[c]{@{}c@{}}-0.385***\\ (0.061)\end{tabular} \\ \hline
\multicolumn{1}{|l|}{Middle author}          & \begin{tabular}[c]{@{}c@{}}0.138**\\ (0.046)\end{tabular}   \\ \hline
\multicolumn{2}{|l|}{\textit{Fixed Effects}}                                                               \\ \hline
\multicolumn{1}{|l|}{Publication Year}       & Y                                                           \\ \hline
\multicolumn{1}{|l|}{Fields}                 & Y                                                           \\ \hline
\multicolumn{1}{|l|}{Country}                & Y                                                           \\ \hline
\multicolumn{2}{|l|}{}                                                                                     \\ \hline
\multicolumn{1}{|l|}{\#Observations}         & 17634                                                       \\ \hline
\multicolumn{1}{|l|}{Log-likelihood}         & -15567.73                                                   \\ \hline
\multicolumn{1}{|l|}{Prob \textgreater $Chi^2$} & \textit{\textless{}0.001}                                   \\ \hline
\end{tabular}

\caption*{** $< 0.01$; *** $< 0.001$}
\label{mlogit}
\end{table}
Table~\ref{mlogit} is the multinomial logit estimate comparing the number of female authors to male authors for the last author and middle author positions relative to the first author position.
The multinomial logit for female authors relative to male authors is $0.14$ times higher for being in the middle position relative to the first position ($p<0.01$). Similarly, the number of female authors is 0$.38$ unit lower than the number of male authors for being in the last position relative to the first position.

\section{Conclusion}
\label{discussion}
Gender disparity in science is widely studied by researchers worldwide; however, the existence of gender disparity in the retracted publication is not investigated yet. This study investigated women's participation in various authorship positions and their likelihood to be in that position.  Also, the proportion of male-female collaboration in such a retracted paper is being analyzed. Out of 3750 retracted papers, the gender of 19164 authors was identified. Overall females acquire 26.43\% of authorship in retracted papers and the share of male-female collaborative retracted publications is 55.11\%. In order to investigate the female contribution at various authorship positions: first, last and middle author, the odds ratio was calculated.  It was observed that women are less likely to hold the last authorship (OR = 0.62,  $p<0.001$) and more likely to hold the middle authorship position (OR= 1.38,  $p<0.001$).

Although we studied the largest set of retracted papers to date, our work is not without limitations. First, for retracted articles published before 2006, data extraction is not feasible, since the author names were predominantly abbreviated with initial letters, making first-name-based gender determination impracticable. Our findings are solely based on the subset of the whole retracted database. Second, we did not compare the study with the controlled group; however, we showed the impact of authorship position with respect to the controlled variable like the publication year, filed, and country in Table ~\ref{mlogit}. Third, we have assumed a first author as the main author and the last as a lead author, which may not always hold true. Fourth, the use of Gender API to identify gender and selection of cutoff. A higher threshold will reduce the count of the number of authors. This study can be further extended to understand the collaboration patterns among males and females.

Our results highlight women's participation and roles in scientific misconduct. First, it shows the important timeline of female involvement in the retracted publication which shows a positive trend. Second, more than 50\% papers retracted were written in collaboration among males and females. Third, females in overall more likely to participate as middle authors than the first and last. This shows less participation of women in scientific misconduct.
\section*{Acknowledgment}

We acknowledge BML Munjal University, India for financial support to buy subscription of Gender-API and to Jawaharlal Nehru University, India for the data access from Web of Science.

 \section*{Conflict of interest}
 The author declares no conflict of interest.
\bibliographystyle{cas-model2-names}

\bibliography{sn-bibliography}

\begin{thebibliography}{12}
\expandafter\ifx\csname natexlab\endcsname\relax\def\natexlab#1{#1}\fi
\providecommand{\url}[1]{\texttt{#1}}
\providecommand{\href}[2]{#2}
\providecommand{\path}[1]{#1}
\providecommand{\DOIprefix}{doi:}
\providecommand{\ArXivprefix}{arXiv:}
\providecommand{\URLprefix}{URL: }
\providecommand{\Pubmedprefix}{pmid:}
\providecommand{\doi}[1]{\href{http://dx.doi.org/#1}{\path{#1}}}
\providecommand{\Pubmed}[1]{\href{pmid:#1}{\path{#1}}}
\providecommand{\bibinfo}[2]{#2}
\ifx\xfnm\relax \def\xfnm[#1]{\unskip,\space#1}\fi
\bibitem[{Agogino(2007)}]{agogino2007beyond}
\bibinfo{author}{Agogino, A.}, \bibinfo{year}{2007}.
\newblock \bibinfo{title}{Beyond bias and barriers: Fulfilling the potential of
  women in academic science and engineering}, in: \bibinfo{booktitle}{APS April
  Meeting Abstracts}, pp. \bibinfo{pages}{K6--001}.
\bibitem[{Aiston and Jung(2016)}]{aiston2016women}
\bibinfo{author}{Aiston, S.J.}, \bibinfo{author}{Jung, J.},
  \bibinfo{year}{2016}.
\newblock \bibinfo{title}{Women academics and research productivity: An
  international comparison}, in: \bibinfo{booktitle}{Globalised re/gendering of
  the academy and leadership}. \bibinfo{publisher}{Routledge}, pp.
  \bibinfo{pages}{17--32}.
\bibitem[{Azoulay et~al.(2017)Azoulay, Bonatti and Krieger}]{azoulay2017career}
\bibinfo{author}{Azoulay, P.}, \bibinfo{author}{Bonatti, A.},
  \bibinfo{author}{Krieger, J.L.}, \bibinfo{year}{2017}.
\newblock \bibinfo{title}{The career effects of scandal: Evidence from
  scientific retractions}.
\newblock \bibinfo{journal}{Research Policy} \bibinfo{volume}{46},
  \bibinfo{pages}{1552--1569}.
\bibitem[{Fang et~al.(2013)Fang, Bennett and Casadevall}]{fang2013males}
\bibinfo{author}{Fang, F.C.}, \bibinfo{author}{Bennett, J.W.},
  \bibinfo{author}{Casadevall, A.}, \bibinfo{year}{2013}.
\newblock \bibinfo{title}{Males are overrepresented among life science
  researchers committing scientific misconduct}.
\newblock \bibinfo{journal}{MBio} \bibinfo{volume}{4},
  \bibinfo{pages}{e00640--12}.
\bibitem[{Fang et~al.(2012)Fang, Steen and Casadevall}]{fang2012misconduct}
\bibinfo{author}{Fang, F.C.}, \bibinfo{author}{Steen, R.G.},
  \bibinfo{author}{Casadevall, A.}, \bibinfo{year}{2012}.
\newblock \bibinfo{title}{Misconduct accounts for the majority of retracted
  scientific publications}.
\newblock \bibinfo{journal}{Proceedings of the National Academy of Sciences}
  \bibinfo{volume}{109}, \bibinfo{pages}{17028--17033}.
\bibitem[{Peng et~al.(2022)Peng, Romero and Horv{\'a}t}]{peng2022dynamics}
\bibinfo{author}{Peng, H.}, \bibinfo{author}{Romero, D.M.},
  \bibinfo{author}{Horv{\'a}t, E.{\'A}.}, \bibinfo{year}{2022}.
\newblock \bibinfo{title}{Dynamics of cross-platform attention to retracted
  papers}.
\newblock \bibinfo{journal}{Proceedings of the National Academy of Sciences}
  \bibinfo{volume}{119}, \bibinfo{pages}{e2119086119}.
\bibitem[{Sharma(2021)}]{sharma2021team}
\bibinfo{author}{Sharma, K.}, \bibinfo{year}{2021}.
\newblock \bibinfo{title}{Team size and retracted citations reveal the patterns
  of retractions from 1981 to 2020}.
\newblock \bibinfo{journal}{Scientometrics} \bibinfo{volume}{126},
  \bibinfo{pages}{8363--8374}.
\bibitem[{Steen(2011)}]{steen2011retractions}
\bibinfo{author}{Steen, R.G.}, \bibinfo{year}{2011}.
\newblock \bibinfo{title}{Retractions in the scientific literature: is the
  incidence of research fraud increasing?}
\newblock \bibinfo{journal}{Journal of medical ethics} \bibinfo{volume}{37},
  \bibinfo{pages}{249--253}.
\bibitem[{Steen et~al.(2013)Steen, Casadevall and Fang}]{steen2013has}
\bibinfo{author}{Steen, R.G.}, \bibinfo{author}{Casadevall, A.},
  \bibinfo{author}{Fang, F.C.}, \bibinfo{year}{2013}.
\newblock \bibinfo{title}{Why has the number of scientific retractions
  increased?}
\newblock \bibinfo{journal}{PloS one} \bibinfo{volume}{8},
  \bibinfo{pages}{e68397}.
\bibitem[{Van~Noorden et~al.(2011)}]{van2011science}
\bibinfo{author}{Van~Noorden, R.}, et~al., \bibinfo{year}{2011}.
\newblock \bibinfo{title}{Science publishing: The trouble with retractions}.
\newblock \bibinfo{journal}{Nature} \bibinfo{volume}{478},
  \bibinfo{pages}{26--28}.
\bibitem[{V{\'a}s{\'a}rhelyi et~al.(2021)V{\'a}s{\'a}rhelyi, Zakhlebin,
  Milojevi{\'c} and Horv{\'a}t}]{vasarhelyi2021gender}
\bibinfo{author}{V{\'a}s{\'a}rhelyi, O.}, \bibinfo{author}{Zakhlebin, I.},
  \bibinfo{author}{Milojevi{\'c}, S.}, \bibinfo{author}{Horv{\'a}t, E.{\'A}.},
  \bibinfo{year}{2021}.
\newblock \bibinfo{title}{Gender inequities in the online dissemination of
  scholars’ work}.
\newblock \bibinfo{journal}{Proceedings of the National Academy of Sciences}
  \bibinfo{volume}{118}, \bibinfo{pages}{e2102945118}.
\bibitem[{Yang et~al.(2022)Yang, Tian, Woodruff, Jones and
  Uzzi}]{yang2022gender}
\bibinfo{author}{Yang, Y.}, \bibinfo{author}{Tian, T.Y.},
  \bibinfo{author}{Woodruff, T.K.}, \bibinfo{author}{Jones, B.F.},
  \bibinfo{author}{Uzzi, B.}, \bibinfo{year}{2022}.
\newblock \bibinfo{title}{Gender-diverse teams produce more novel and
  higher-impact scientific ideas}.
\newblock \bibinfo{journal}{Proceedings of the National Academy of Sciences}
  \bibinfo{volume}{119}, \bibinfo{pages}{e2200841119}.

\end{thebibliography}

\end{document}